\documentclass[pdflatex,sn-chicago]{sn-jnl}% Chicago-based Humanities Reference Style

\usepackage{graphicx}%
\usepackage{multirow}%
\usepackage{amsmath,amssymb,amsfonts}%
\usepackage{amsthm}%
\usepackage{mathrsfs}%
\usepackage[title]{appendix}%
\usepackage{xcolor}%
\usepackage{textcomp}%
\usepackage{manyfoot}%
\usepackage{booktabs}%
\usepackage{algorithm}%
\usepackage{algorithmicx}%
\usepackage{algpseudocode}%
\usepackage{listings}%
\usepackage{siunitx}
\usepackage{array}
\usepackage{makecell}

\theoremstyle{thmstyleone}%
\theoremstyle{thmstyletwo}%

\theoremstyle{thmstylethree}%

\begin{document}

\title[Article Title]{QualiTagger: Automating software quality detection in issue trackers}

%%=============================================================%%
%% GivenName	-> \fnm{Joergen W.}
%% Particle	-> \spfx{van der} -> surname prefix
%% FamilyName	-> \sur{Ploeg}
%% Suffix	-> \sfx{IV}
%% \author*[1,2]{\fnm{Joergen W.} \spfx{van der} \sur{Ploeg} 
%%  \sfx{IV}}\email{iauthor@gmail.com}
%%=============================================================%%

\author*[1]{\fnm{Karthik} \sur{Shivashankar}}\email{karths@ifi.uio.no}

%\equalcont{These authors contributed equally to this work.}

\author[2]{\fnm{Rafael} \sur{Capilla}}\email{rafael.capilla@urjc.es}

\author[3]{\fnm{Maren} \sur{Maritsdatter Kruke}}\email{maren.kruke@visma.com}

\author[3]{\fnm{Mili} \sur{Orucevic}}\email{mili.orucevic@visma.com}
%\equalcont{These authors contributed equally to this work.}
\author[1]{\fnm{Antonio} \sur{Martini}}\email{antonio.martini@ifi.uio.no}

\affil*[1]{\orgdiv{Department of Informatics}, \orgname{University of Oslo}, \city{Oslo}, \country{Norway}}

\affil[2]{\orgdiv{Department of Informatics}, \orgname{Rey Juan Carlos University}, \orgaddress{\street{Tulipán s/n}, \city{Móstoles}, \postcode{28933}, \state{Madrid}, \country{Spain}}}

\affil*[3]{\orgname{Visma Software International}, \city{Oslo}, \country{Norway}}

%%==================================%%
%% Sample for unstructured abstract %%
%%==================================%%

\abstract{A systems quality is a major concern for development teams when it evolve. Understanding the effects of a loss of quality in the codebase is crucial to avoid side effects like the appearance of technical debt. Although the identification of these qualities in software requirements described in natural language has been investigated, most of the results are often not applicable in practice, and rely on having been validated on small datasets and limited amount of projects. For many years, machine learning (ML) techniques have been proved as a valid technique to identify and tag terms described in natural language. In order to advance previous works, in this research we use cutting edge models like Transformers, together with a vast dataset mined and curated from GitHub, to identify what text is usually associated with different quality properties. We also study the distribution of such qualities in issue trackers from openly accessible software repositories, and we evaluate our approach both with students from a software engineering course and with its application to recognize security labels in industry.
}


%%\pacs[JEL Classification]{D8, H51}

%%\pacs[MSC Classification]{35A01, 65L10, 65L12, 65L20, 65L70}

\maketitle

\section{Introduction}\label{intro}
Enhancing system quality is a primary objective for software organizations, aimed at mitigating adverse user impacts and minimizing the accrual of technical debt.
 
Code analyzers, such as SonarQube\footnote{https://www.sonarsource.com/}, often highlight the specific quality attributes (e.g. reliability, maintainability, etc.) most impacted by detected issues. This allows quality engineers to prioritize and address technical debt strategically and align with their organization's goals.

Nevertheless, while such tools are able to detect specific parts of the code affected by code and design flaws, there are many other issues affecting the quality of systems that cannot be automatically recognized by static analysis tools and automatic techniques \citep{martini2018}. Let's take, for instance, an optimal architectural solution (e.g. the choice of a database) that is soon becoming sub-optimal because of the need to implement a new feature that requires different performance requirements (e.g. a new database). The loss of performance may depend on the new future requirement and cannot, therefore, be recognized by a static code analyzer in the current code as suboptimal. In such cases, developers, quality engineers and architects usually create issues in issue trackers such as Jira or GitHub to signal that, before implementing the new feature, the performance needs to be improved (in the example, the database needs to be substituted). Good practices suggest that such issues should be tagged with the name of the affected quality (in our example, the database issue would be tagged with "performance"). When the backlog is groomed, the issues that are related to strategically important qualities for the organization, can be prioritized for the upcoming iteration. In addition, such tags allow the organization to obtain useful insights, for example on the number of issues that are currently degrading one specific quality. This can, in turn, show if the architectural trade offs are suboptimal.

However, as discussed in \citep{Cabot_2015}, most of the practitioners do not tag their tasks and requirements in open source software. The authors \citep{Cabot_2015} report that, out of 3,757,038 studied projects, only 122,012 (3.25\%) of them contain labeled issues, despite a positive correlation showing that the issues that are being labeled are also more frequently solved. This indicates that labeling issues might be beneficial, but it is not done extensively because tagging implies an overhead for developers.

Automatically tagging issues would make labels more accessible in projects without burdening developers. Also, the increasing availability of data to create a vast dataset together with the continuous improvement of Deep Learning (DL) models like transformers.  Nowadays, there are almost no studies dedicated to classifying issues with labels related to the affected qualities, which highlight a substantial gap in this direction. On the other hand, a number of studies have looked into how to automatically classify non-functional requirements (e.g. \citep{Fazelnia2024}): such studies, although focused on general requirements and not on issue trackers and system qualities, are relevant to understand the state of the art. In most cases, such studies provide good results in using NLP, and even transformers, to classify NFR. However, in most cases, they are based on training and validating NLP models on relatively small datasets of requirements (and not issue trackers) and do not provide large datasets that would make the trained models more powerful and generalizable. In addition, the results are often only validated on issues from the same projects that the models were trained on, without testing the performance on projects that have not been seen by the model during training (Out of Distribution projects, OOD). Finally, with a few exceptions, such results are still not very performing in terms of precision and recall, as they use outdated architecture and models like Recurrent Neural Network (RNN) and Convolution Neural Network (CNN)  (predating transformers), which limits their application in practice. 

In this study, we aimed at improving the state of the art by developing a new model, \textbf{QualiTagger}, based on Transformers architecture. The intent is to analyze text in issues from issue trackers (e.g. Jira and GitHhub) to automatically identify if the reported descriptions lead to understanding if they influence specific system qualities. This will give additional information to the organization on how to prioritize issues or on which qualities are more at risk. We also provide a novel curated and vast dataset (i.e. QualiDataSet) containing millions of issues labelled with different qualities mined from thousands of GitHub projects.

In particular, in this study, we aim to answer the following questions:

\textit{\textbf{RQ1: How effective are Transformer models at classifying software quality attributes in issue tracking systems?}}

\textit{\textbf{RQ1.1: To what extent are Transformer models able to generalize their quality classification when applied to previously unseen projects?}}

\textbf{Rationale:} RQ1 aims to investigate the capabilities of transformer models to accurately classify software quality concerns within issue trackers.  We aim to determine if transformers can effectively analyze textual descriptions of issues and assign them to appropriate software quality categories (e.g., performance, security, usability). We aim at evaluating the use of the trained models not only on the test split of the dataset, but also on projects that have not been used for training, and therefore are unseen ("Out of Distribution" projects)  by the models. This would show us to what extent our models can be used out of the box to classify issues in new projects.

\textit{\textbf{RQ2: Which classification strategy performs better for software quality identification: a multiclass approach or multiple binary classifiers?}}

\textbf{Rationale:} {In this research question, we explore the comparative effectiveness of two distinct classification approaches to identify software quality concerns in issue trackers. We investigate whether a single multiclass classifier, which assigns an issue to one of multiple quality categories simultaneously,  outperforms (or not) a multiple binary classifiers strategy, where each classifier is trained to detect the presence or absence of a specific quality. This comparison will provide insights into the optimal classification architecture for this task.}

\textit{\textbf{RQ3: Is an LLM or a multiple binary classifier better at identifying qualities in issues?}}

\textbf{Rationale:} {With RQ3 we aim to understand if a focused approach, such as a set of small and specialized transformer models, is more or less effective than general and vast models such as GPT 4o model. This has implications in terms of effectiveness and accessibility, but also in terms of resources and energy efficiency.}

To answer the aforementioned questions, we provide the following contributions:

\begin{itemize}
  \item We release a vast and curated dataset, \textbf{QualiDataSet}, containing issues tagged with seven different qualities. Such dataset can help creating new models and conduct large-scale studies related to these quality attributes.  
  \item We create an approach, \textbf{QualiTagger}, encompassing an ensemble of binary transformers models, that can reliably tag text according to the qualities it is related to. We assess the  accuracy of such an approach, and we confirm its superior performance with respect to both multiclass classifiers and powerful LLMs. We assess the usefulness in practice of our approach with students in a course project, and its very promising performance when applied in industrial projects.
  \item We showcase how our model can identify and analyze quality concerns in projects and datasets related to specific domains and software repositories. In particular, we analyze thousands of Technical Debt (TD) issues, to understand if TD is related to distinct qualities in a specific project or in a specific programming language.
\end{itemize}

\section{Background}\label{background}
In this section we describe some background and related works regarding the detection of quality issues in software projects. We also discuss ML techniques used for quality detection and the role of NLP transformers.

\subsection{Software Quality and its Challenges}

The identification of quality issues in software projects is crucial due to their impact on maintainability and evolution \citep{Curtis2022}. These issues can manifest as structural code defects \citep{Curtis2022}, violations of quality attributes \citep{Rahimi2020}, or even architectural weaknesses \citep{Curtis2022}. Traditional quality detection tools primarily rely on static code analysis \citep{Avgeriou2021, Curtis2012}, which may overlook quality issues arising from evolving requirements or future system changes. Furthermore, the diverse range of quality models and metrics employed by these tools can lead to inconsistent and incomparable quality assessments across different projects \citep{Letouzey2016, Avgeriou2021, Curtis2012}. This highlights the need for more robust and adaptable approaches to software quality detection.

\subsection{Transformers and their Advantages}

Transformers have revolutionized Natural Language Processing (NLP) with their ability to capture contextual dependencies in text through attention mechanisms \citep{NIPS2017_3f5ee243}. This allows for improved parallelization and faster training times. The Hugging Face (HF) library \citep{wolf2019huggingface} has further facilitated the adoption of transformers by providing a repository of pre-trained models and accessible APIs for various architectures like BERT \citep{devlin2019bert}, and GPT \citep{brown2020language}. These models can be fine-tuned for specific tasks, empowering researchers to achieve state-of-the-art results even with limited labeled data.

\subsection{Machine Learning for Software Quality Detection}
Machine learning (ML) has emerged as a powerful tool for predicting and estimating software quality, yet its application is far from seamless. While recent studies have demonstrated the potential of ML in various aspects of quality detection, a critical examination reveals several persistent challenges and opportunities for advancement. The application of ML to software quality is not a simple matter of adopting existing algorithms; it requires a nuanced understanding of the unique characteristics of software development and the inherent complexities of quality attributes.

One prominent area of research is the \textbf{classification of Non-Functional Requirements (NFRs)}. Studies employing Convolutional Neural Networks (CNNs) \citep{Cody2019} and traditional supervised ML algorithms like Logistic Regression \citep{Khatian2021} have showcased promising results. However, a significant limitation lies in the robustness and generalizability of these models. The ability to accurately classify NFRs across diverse project contexts and datasets remains a challenge. The inherent ambiguity and subjectivity in NFR definitions, coupled with the variability in project domains, often lead to inconsistencies in classification accuracy. Furthermore, the lack of extensive, high-quality labeled datasets hinders the training of robust models, particularly for less common NFR types.

Moving beyond individual requirements, the \textbf{prediction of system quality} is another critical area. Surveys on ML and deep learning techniques \citep{ALASWAD2022} have highlighted the focus on attributes like reliability, maintainability, and reusability. However, a critical gap exists in the integration of contextual factors, such as development processes, team dynamics, and organizational culture, into predictive models. Current methods often rely heavily on code metrics and historical data, neglecting the broader organizational context that significantly influences software quality. The absence of standardized evaluation benchmarks and the difficulty in obtaining high-quality labeled data further complicate the development and comparison of effective prediction models.

Finally, \textbf{detecting code smells} using ML algorithms like Decision Trees, XGBoost, and SVM \citep{Khleel2023} has shown considerable success. Yet, there is a pressing need for more fine-grained detection and classification of complex code smells. Current methods often struggle with identifying subtle and context-dependent smells, and the interpretability of ML-based detection results remains a challenge. Integrating ML-based code smell detection with automated refactoring techniques is a promising avenue for future research, but the practical realization of this integration requires overcoming several technical and conceptual hurdles.

\subsection{Transformer-based Classification in Software Quality Detection}

Transformer-based models have revolutionized various natural language processing tasks, and their application in software engineering, particularly in classification, has yielded remarkable results. However, the adoption of these models is not without its challenges and limitations, necessitating a critical evaluation and a focus on future directions.

In the domain of \textbf{NFR classification}, while combining improved BERT models with Bi-LSTM networks \citep{Li2022} and leveraging BERT's transfer learning capabilities \citep{Chatterjee2021} have demonstrated significant improvements, the computational cost and resource requirements of these models remain substantial. The dependence on large labeled datasets also restricts their applicability in domains with limited labeled data. Although Natural Language Inference (NLI) pipelines have shown superior performance \citep{Fazelnia2024}, the interpretability of transformer-based NFR classification models remains a critical concern, hindering their practical adoption in safety-critical and regulated domains. A crucial area for future research is the development of techniques to adapt pre-trained transformer models to the specific nuances of NFR language.

Similarly, in \textbf{classifying technical debt}, while NLP methods, particularly transformer-based models, have achieved success in automatically classifying technical debt in issue trackers \citep{skryseth2023}, accurately capturing the semantic nuances and context of technical debt descriptions remains a significant challenge. The classification of technical debt often requires understanding complex dependencies and trade-offs, which are difficult for current models to learn. The lack of standardized taxonomies and labeled datasets for technical debt classification further impedes the development of robust and generalizable models.

In a broader context, recent reviews have emphasized the importance of transformers in requirements classification \citep{Sonbol2022, Kaur2024}, highlighting the effectiveness of transfer learning techniques in BERT \citep{devlin2019bert} and DistillBERT. However, a major limitation lies in the evaluation of these models in real-world software development contexts. The transfer learning capabilities of these models need to be further explored for domain adaptation and cross-project learning, especially in scenarios with limited data. Addressing the interpretability and explainability of transformer-based models is crucial for building trust and facilitating their adoption in software quality detection.

In particular, our work improves previous state of the art according to the following points:
\paragraph{1. Contrast with NFR Classification:}

Correctly and automatically tagging issues would make labels more accessible in projects without burdening developers. There are almost no studies dedicated to classifying the qualities affected by issues in issue trackers, which highlights a substantial gap in this direction.

On the other hand, a number of studies have looked into how to automatically classify Non-Functional Requirements (e.g. \citep{Fazelnia2024}): such studies, although limited to requirements, are relevant to learn what has been working and not. In most cases, such studies provide good results in using NLP, and even transformers, to classify NFR. However, these studies primarily focus on classifying requirements documents, not the less structured and more diverse language found in issue trackers. Furthermore, in most cases they are based on training and validating NLP models on relatively small datasets of requirements (and not issue trackers) and do not provide large datasets that would make the trained models more powerful and generalizable.

\paragraph{2. Generalizability of results and evaluation in Out of Distribution projects:}

In addition, the results are often only validated on issues from the same projects that the models were trained on, without testing the performance on projects that have not been seen by the model during training (Out of Distribution projects, OOD). This limitation restricts the practical applicability of existing models, as they may not generalize well to new and unseen projects.

\section{Research design}\label{method}

Our research methodology follows a case study framework, investigating the automatic classification of software quality attributes from GitHub issues. We adhered to the Empirical standards for conducting and evaluating research in software engineering \citep{sigsoft_empirical_standards_nodate}.

Our work comprises two main components: an evaluative case study assessing QualiTagger's effectiveness, and an exploratory case study examining different software project issues in Github to investigate the relationship between Technical Debt and software quality attributes across programming languages from  Ground Truth Technical Debt (GTD) \citep{skryseth2023} Dataset. Our methodology incorporates both quantitative data through model performance metrics and qualitative data via student feedback, enabling methodological triangulation to strengthen our findings.

\subsection{Case Study Methodology: Rigorous Evaluation and Real-World Application}

Our case study implemented a structured protocol to evaluate automated software quality attribute classification. Initially, we curated a balanced dataset from GitHub issues, ensuring diverse representation of quality attributes. 

The empirical core involved training and testing machine learning models using cross-validation across various projects for robust validation. Beyond numerical metrics, we focused on interpreting classification complexities, establishing a clear evidence chain. 

This research prioritized real-world context by utilizing authentic GitHub data, addressing practical challenges in issue classification. Direct observations through student evaluations provided qualitative insights into the tool's practical application, comparing manual and automated tagging.

Exploratory technical analysis of Technical Debt's impact across programming languages revealed significant patterns in software quality practices. We ensured research transparency and reproducibility through meticulous documentation, including a comprehensive replication package detailing data preprocessing, model architectures, and evaluation metrics. This methodological rigor, aligned with case study standards, delivers meaningful insights into automated software quality attribute classification in issue tracking systems.

\begin{figure*}
    \centering
    \includegraphics[width=1\linewidth]{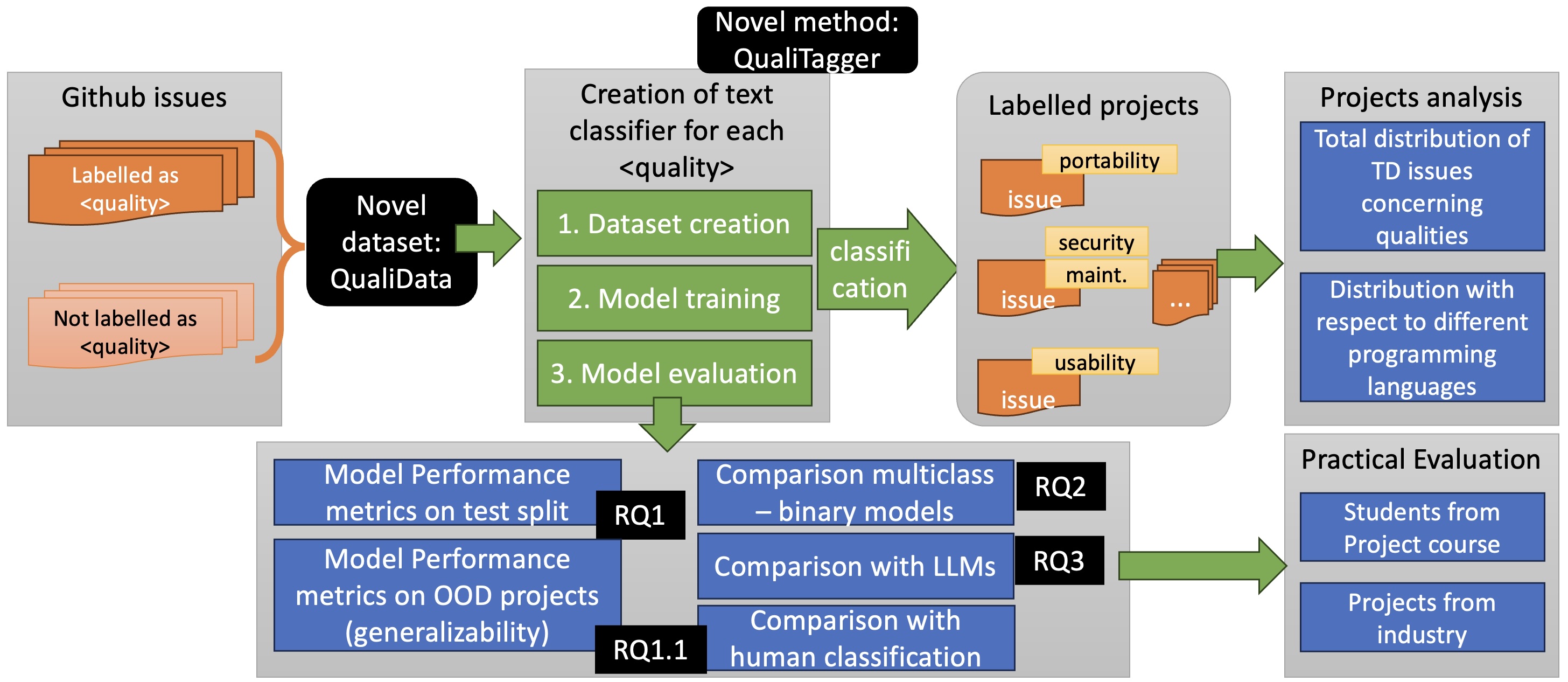}
    \caption{Research Process showing sources, steps, results related to various RQs, novel contribitions (black boxes) and finally both usage showcase of our tool and evaluation in practice}
    \label{fig:design}
\end{figure*}

%\section{Research Methods}\label{approach}
Our research design is reported in Fig. \ref{fig:design}. The first goal to answer our RQs is to classify text that concerns specific qualities automatically. We used DistilRoBERTa \citep{Sanh2019DistilBERTAD}, a compact version of the RoBERTa model, to classify text related to specific software qualities. These qualities – Maintainability, Security, Reliability, Usability, Compatibility, Performance, and Portability – were selected based on the SQuaRE ISO standard.

The {\textbf{\textit{Replication Package can be found at this Link.}}}\footnote{https://zenodo.org/doi/10.5281/zenodo.10148503}

\subsection{Dataset creation}
To develop an efficient binary classification model using GitHub issues, we used a targeted data mining approach. We extracted the data from January 2015 to September 2023 \citep{GHArchive}, focusing on issues explicitly labeled with terms that align with the seven key software quality attributes identified in the SQuaRE series. These attributes include \textbf{Maintainability, Security, Reliability, Usability, Compatibility, Performance, and Portability} qualities. Issues marked with these labels represent positive instances for each attribute's respective dataset. In particular, we used the following regular expressions to select the labels:\\

\textit{maintain*,  securit*,  reliab*,  usab*, compatib*,  scal*,  perform*, portab*}\\

where each '*' means that the initial string can be followed by any suffix except new line.

\subsection{Dataset Processing and Cleaning}

The dataset obtained from GHArchive \citep{GHArchive} required extensive preprocessing to ensure optimal model training conditions. Our cleaning pipeline addressed several key challenges in the raw data:

\begin{itemize}
    \item \textbf{Duplicate Elimination}: We removed duplicate issues to prevent model bias and ensure each training instance contributed unique information.
    
    \item \textbf{Language Processing}: 
        \begin{itemize}
            \item Converted all text to lowercase for consistency
            \item Removed non-English text to maintain linguistic uniformity
            \item Eliminated URLs, emojis, and special characters that could introduce noise
        \end{itemize}
    
    \item \textbf{Content Validation}: We established minimum character thresholds for issues, removing those too brief to provide meaningful classification signals.
\end{itemize}

For the binary classification task, we implemented a strategic approach to construct the non-positive class. For each quality attribute under investigation, we created negative samples by randomly selecting issues that were explicitly labeled with other quality attributes from our set of seven qualities. This methodology ensured that our negative class represented a diverse range of alternative quality concerns, strengthening the model's discriminative capabilities.

To prevent class imbalance, which could bias the model's predictions, we carefully balanced our training data. For each quality attribute shown in Table \ref{tab:quality-type-distribution}, we maintained a 1:1 ratio between positive and negative examples. This balanced dataset structure was crucial for training our DistilRoBERTa model, ensuring it could effectively, distinguish between the presence and absence of specific quality attributes and generate reliable predictions across diverse software projects.

This rigorous preprocessing pipeline established a robust foundation for our model to learn meaningful patterns across different quality types.

\begin{table}[htbp]
    \centering
    \caption{Count of Positive Quality Type }
    \label{tab:quality-type-distribution}
    \begin{tabular}{lr}
        \toprule
        \textbf{Quality Type} & \textbf{Count of Positive Instance}\\
        \midrule
        Security & 277,806 \\
        Performance & 46,413 \\
        Compatibility & 17,521 \\
        Maintainability & 4,410 \\
        Portability & 1,406 \\
        Reliability & 1,745 \\
        Usability & 21,287 \\
        \bottomrule
    \end{tabular}
\end{table}

\subsection{Model training}

\subsubsection{Model Architecture and Training Methodology}
We employed DistilRoBERTa \citep{Sanh2019DistilBERTAD} as our base model, selected for its efficiency and reduced complexity compared to RoBERTa while maintaining comparable performance. The model underwent fine-tuning for each software quality attribute to adapt its pre-trained knowledge to our specific domain context. This fine-tuning process was crucial for interpreting domain-specific language in software development issues, understanding the contextual nuances of developer communications, and distinguishing between different software quality attributes in natural language.

For individual quality types ( binary classifier appraoch), we employed 5-fold cross-validation on balanced datasets to prevent overfitting, ensure robust performance metrics, and enable reliable model evaluation.

In the multiclass classifier, we utilized stratified 5-fold cross-validation over five epochs, incorporating class-weighted loss functions to address class imbalance. This approach maintained proportional representation of classes across folds \citep{Kfold}. The class weights were dynamically assigned based on class frequency, ensuring adequate learning from underrepresented classes \citep{weighteclass}.

\subsubsection{Technical Implementation and Processing}
Our training configuration was optimized with specific hyperparameters: training batch size of 32, evaluation batch size of 64, and a cosine scheduler learning rate with 500 warm-up steps. We implemented weight decay of 0.01 and CrossEntropyLoss as the loss function. The implementation utilized FP16 precision for computational efficiency, with gradient accumulation over 2 steps and model checkpointing retaining the latest version.

The data processing pipeline centered on text preprocessing through tokenization with consistent maximum sequence length, implementing both padding and truncation as needed. This standardized approach ensured uniform input structure while preserving essential semantic information. The binary classification approach proved particularly effective, allowing for enhanced focus on individual quality attributes and demonstrating reduced sensitivity to class imbalance compared to multiclass approaches. The simplified decision boundaries of binary classifier led to more reliable classification, particularly beneficial for identifying less frequent quality types, enabling our model to maintain high accuracy across all quality attributes, regardless of their frequency in the dataset compared to multiclass approach.

\subsection{Model Evaluation}

The evaluation phase is designed to rigorously test the model's ability to apply its learned knowledge to classify new, unseen data. For this project, testing involved presenting the fine-tuned DistilRoBERTa models with a set of GitHub issues that it had not encountered during the training phase. This is a crucial step in the machine learning workflow, as it simulates the model's real-world performance and uncovers its ability to generalize from its training data to other data.

During testing, each issue in the test dataset was processed by the model, which then had to classify it as either positive or non-positive for the specific software quality attribute it was trained on. The confidence threshold was set at an exacting 0.9 to reduce the incidence of false positives. This threshold implies that the model is highly certain of its classifications, which is particularly important in applications where the cost of misclassification is high.

The testing process not only evaluates the model's predictive accuracy but also allows for the refinement of the confidence threshold by observing its impact on the balance between precision and recall — crucial metrics in the context of binary classification problems.

As an additional step, we have tested our models by classifying one Out of Distribution (OOD)  project for each quality. This way, we also tested how much would the model generalize or, in other words, recognize issues that do not come from the same context of the dataset. 

For each quality attribute, we employed a \textbf{leave-one-out} strategy, where we excluded the project with the highest number of issues for that specific quality from the training set.

For instance, in the case of \textit{"maintainability,}" we removed the \textit{'ansible'} project, which had the most issues tagged by developers as related to maintainability (or terms matching the corresponding regex). We then trained the model on this reduced dataset and subsequently tested the classifier on the excluded project. This process was repeated for all quality attributes.

This methodology served two purposes:
\begin{enumerate}
    \item  It mitigated potential bias from larger projects dominating the classifier's training.
    \item  It allowed us to assess the model's ability to generalize to unseen data.
\end{enumerate}

This evaluation was particularly crucial for our exploratory case study scenario, where we needed to classify issues tagged as Technical Debt (TD) in the Ground Truth Technical Debt  (GTD) Dataset \citep{skryseth2023} that lacked specific quality attribute tags. As a result, we could not directly test the performance against available ground truth for these issues.

These results suggest that our approach is robust and capable of identifying quality-related issues even in projects outside the training distribution, which is essential for practical application in diverse software engineering contexts.

\subsection{Evaluation metrics}
We employed comprehensive evaluation of the trained models using multiple performance metrics. We employed standard metrics including accuracy, precision, recall, Matthews Correlation Coefficient (MCC), Area Under the Receiver Operating Characteristics Curve (AUC-ROC), and F1-score on a\textbf{ held-out test set} and on \textbf{Leave-one-out OOD }projects.

Precision quantifies the proportion of correctly identified quality issues (true positives) among all predicted issues. It is calculated as:

\[
Precision = \frac{True\;Positives}{True\;Positives + False\;Positives}
\]

Recall, a particularly crucial metric in our study, measures the proportion of correctly identified quality issues (true positives) among all actual issues. It is defined as:

\[
Recall = \frac{True\;Positives}{True\;Positives + False\;Negatives}
\]

Recall's significance in our context stems from the nature of software repository data, where quality issues are often only partially tagged. This partial labeling means that our negative class may inadvertently contain unlabeled positive instances. While this affects the reliability of precision (which depends on both positive and negative label accuracy), recall remains robust as it focuses solely on identifying known quality issues.

The F1-score provides a balanced measure by combining precision and recall through their harmonic mean:

\[
F1 = 2 \cdot \frac{Precision \cdot Recall}{Precision + Recall}
\]

The AUC-ROC metric evaluates model performance across various classification thresholds by measuring the area under the curve that plots the true positive rate against the false positive rate. An AUC of 0.5 indicates random performance, while 1.0 represents perfect classification.

We particularly emphasize Matthews Correlation Coefficient (MCC), recognized for its statistical reliability in binary classification \citep{Chicco2023MCC}:

{\scriptsize
\[
MCC = \frac{TP \cdot TN - FP \cdot FN}{\sqrt{(TP + FP) \cdot (TP + FN) \cdot (TN + FP) \cdot (TN + FN)}}
\]
}

MCC provides a more comprehensive evaluation than accuracy or F1-score, as it yields high scores only when a model performs well across all confusion matrix categories, accounting for both positive and negative class sizes. Its values range from -1 to 1, where values above 0 indicate better-than-random performance.

\subsection{Statistical Analysis for Model Comparison}
To validate our model's superiority over the GPT model, we implemented a comprehensive statistical analysis framework using four complementary approaches:

1. \textbf{McNemar's Test}: Our primary statistical test, chosen for its effectiveness in comparing paired nominal data in binary classification. This test specifically analyzes the disagreements between models, providing robust evidence of performance differences \citep{sundjaja2023mcnemar, Supervise_stat}.

2. \textbf{Bootstrap Comparison}: We conducted bootstrap analysis of F1 scores with 95\% Confidence Intervals (CI). This non-parametric method provides robust performance difference estimates without distribution assumptions, offering insights into the stability of performance differences across dataset resamples \citep{JMLR:v7:demsar06a}.

3. \textbf{T-test Analysis}: We performed t-tests on F1 scores to compare mean model performance. While this test assumes normality, it provides a widely understood measure of statistical significance \citep{student1908probable}.

4. \textbf{Cliff's Delta}: We calculated Cliff's Delta for F1 scores to quantify effect size. This metric was selected for its suitability with ordinal data and resistance to outliers \citep{vargha2000critique}.

\section{Results}\label{results}

\subsection{RQ1 - Performance of the model in Recognizing Qualities}
The evaluation of DistilRoBERTa's performance in classifying software quality attributes  is quantified using several statistical metrics, each providing unique insights into the model’s predictive power and reliability, as shown in Table \ref{tab:performance}. The precision metric indicates the proportion of true positive predictions in all positive predictions made by the model. High precision signifies that when the model predicts an issue to be relevant to a particular quality, it is likely correct. Recall, on the other hand, assesses the model's ability to identify all relevant instances within the dataset. In terms of binary classification, a high recall indicates that the model has correctly identified most of the issues that are true examples of a quality attribute.

The F1 score provides a harmonic balance between precision and recall, offering a single measure of the model’s accuracy. It is particularly useful when the cost of false positives and false negatives is equally important. An F1 score approaching 1 suggests a robust model that provides reliable and balanced classification performance \citep{scikit-learn_metrics}. The Matthews Correlation Coefficient (MCC) serves as a more comprehensive metric, taking into account true and false positives and negatives. This coefficient is regarded as a balanced measure, useful even if the classes are of very different sizes \citep{Chicco2023MCC}.

To analyze the results, we examine these metrics across the different quality attributes. A high MCC  value in classifying maintainability issues, for instance, will imply a strong correlation between the model’s predictions and the actual distribution of maintainability-related issues.  

First, we report the performance of the model on the test set (a test split randomly selected among all the issues available). As visible in table \ref{tab:performance}, all the models (one for each quality) performed really well on the test data.

\begin{table}[htbp]
\centering
\setlength{\belowcaptionskip}{10pt}
\caption{Performance of our binary classification models in recognizing Software Qualities in text}
\label{tab:performance}

\begin{tabular}{lccccccc}
\toprule
Quality & Precision &  Recall &  Accuracy &  MCC &  F1 Score &  AUC \\
\midrule
 Main & 0.895 & 0.857 & 0.878 & 0.757 & 0.876 & 0.957 \\
 Secu & 0.957 & 0.979 & 0.967 & 0.935 & 0.968 & 0.995 \\
 Reli & 0.853 & 0.960 & 0.897 & 0.801 & 0.903 & 0.961 \\
 Usab & 0.921 & 0.937 & 0.928 & 0.857 & 0.929 & 0.977 \\
 Comp & 0.932 & 0.934 & 0.933 & 0.866 & 0.933 & 0.979 \\
 Perf & 0.949 & 0.946 & 0.947 & 0.895 & 0.947 & 0.985 \\
 Port & 0.906 & 0.820 & 0.867 & 0.738 & 0.861 & 0.954 \\
\bottomrule
\end{tabular}
\end{table}

\paragraph{Out-of-Distribution Performance Analysis}

We evaluated our models' generalization capabilities through Out-of-Distribution (OOD) testing using a Leave-one-out cross-validation approach on unseen projects, as detailed in Table \ref{tab:unseen}. This evaluation was crucial to assess how well our models would perform when applied to entirely new software repositories and contexts. The results demonstrated robust generalization capabilities:

\begin{itemize}
    \item While the performance metrics were slightly lower compared to our standard test set results (as expected for OOD scenarios), they remained consistently strong across different projects
    \item The models maintained reliable prediction capabilities even when faced with previously unseen development contexts and coding patterns
\end{itemize}

These encouraging OOD results provided strong evidence for our models' practical applicability in two key scenarios:
\begin{enumerate}
    \item Identifying quality issues across diverse software project repositories beyond our training data
    \item Analyzing different quality issues within the GTD \citep{skryseth2023} dataset for our exploratory case study
\end{enumerate}

The demonstrated generalization ability suggests that our approach can be effectively deployed as a reliable tool for quality issue detection across varied software development contexts.

\begin{table}[htbp]
\centering
\setlength{\belowcaptionskip}{5pt}
\caption{Performance of our models in recognizing the qualities on OOD (Out of Distribution) projects (generalization). The unseen projects were selected as the ones having the most issues tagged with such quality out of the whole dataset}
\label{tab:unseen}
\begin{tabular}{@{}lcccccccl@{}}
\toprule
Category & Repository with the most tagged issues & Prec. & Rec. & Acc. & MCC & F1 & AUC \\
\midrule
Maintainability & ansible/ansible-modules-core & 0.914 & 0.992 & 0.949 & 0.902 & 0.951 & 0.995 \\
Security & elastic/kibana & 0.942 & 0.748 & 0.851 & 0.717 & 0.834 & 0.935 \\
Reliability & Azure/azure-sdk-for-net & 0.855 & 0.992 & 0.912 & 0.834 & 0.918 & 0.993 \\
Usability & godotengine/godot & 0.920 & 0.938 & 0.928 & 0.856 & 0.929 & 0.974 \\
Compatibility & pingcap/tidb & 0.927 & 0.823 & 0.879 & 0.763 & 0.872 & 0.959 \\
Performance & flutter/flutter & 0.949 & 0.960 & 0.954 & 0.909 & 0.955 & 0.990 \\
Portability & magnumripper/JohnTheRipper & 0.891 & 0.860 & 0.877 & 0.755 & 0.875 & 0.968 \\
\bottomrule
\end{tabular}
\end{table}

\subsection{RQ2: Comparison between binary and multiclass  classifiers}

In contrast to the binary classification models discussed in RQ1, focused on identifying the presence or absence of a single quality attribute, multiclass classifiers tackle the challenge of assigning each issue to one or more of several predefined quality categories. 

The performance of multiclass classifiers is evaluated using metrics similar to those used in binary classification, but with adaptations for the multi-class context. Precision, recall, and F1-score are calculated for each class individually, providing insight into the model's ability to correctly identify and classify issues within that specific category. Additionally, macro and weighted averages of these metrics offer a broader perspective on the overall performance across all classes.

Comparing Tables \ref{table:multi_class_test} and \ref{table:multi_class_top_repo} to the results from RQ1,  highlights several key differences between the binary and multiclass approaches:

\textit{Performance Variation Across Classes:} Multiclass classifiers often exhibit significant performance disparities between classes. This is evident in the considerably lower precision, recall, and F1-scores for certain categories, such as "Portability" in Table \ref{table:multi_class_top_repo}. This suggests that some quality attributes are inherently more challenging to distinguish from others within a multiclass framework.

\textit{Impact of Class Imbalance:} The imbalanced distribution of classes in the dataset can significantly influence multiclass classifier performance. Classes with fewer examples, like "Portability" and "Reliability," tend to suffer from lower performance due to the model's limited exposure to these categories during training.

\textit{Trade-offs Between Precision and Recall: }Multiclass classifiers often face trade-offs between precision and recall. A model might achieve high precision for one class at the expense of lower recall for another. This balance depends on the specific application and the relative importance of minimizing false positives versus false negatives.

Complexity and Interpretability: Multiclass classifiers can be more complex and challenging to interpret than binary models. Understanding the factors influencing the model's decisions for each class can be less straightforward, potentially hindering the identification of areas for improvement.

Overall, the comparison between binary and multiclass classification reveals the inherent complexities and challenges associated with categorizing software quality attributes. While binary classifiers excel at identifying the presence or absence of a single quality, multiclass models offer a more direct categorization approach but grapple with class imbalances, performance variations, and interpretability concerns. The choice between these approaches depends on the specific requirements of the software development process and the trade-offs between accuracy, efficiency, and interpretability.

\begin{table}[ht]
\centering
\caption{Classification Report for Multi-class Test}
\begin{tabular}{lcccc}
\hline
Class & Precision & Recall & F1-Score & Support \\
\hline
Maintainability & 0.647 & 0.559 & 0.600 & 578 \\
Portability & 0.774 & 0.562 & 0.651 & 201 \\
Performance & 0.900 & 0.906 & 0.903 & 6757 \\
Reliability & 0.559 & 0.427 & 0.484 & 234 \\
Security & 0.984 & 0.988 & 0.986 & 41197 \\
Usability & 0.789 & 0.781 & 0.785 & 2955 \\
Compatibility & 0.800 & 0.805 & 0.803 & 2571 \\
\hline
Accuracy & \multicolumn{4}{c}{0.949} \\
Macro Avg & 0.779 & 0.718 & 0.745 & 54493 \\
Weighted Avg & 0.948 & 0.949 & 0.948 & 54493 \\
\hline
\end{tabular}

\label{table:multi_class_test}
\end{table}
\begin{table}[h]

\centering
\caption{Classification Report for Multi-class on OOD projects}
\begin{tabular}{lcccc}
\hline
Class & Precision & Recall & F1-Score & Support \\
\hline
Maintainability & 0.871 & 0.945 & 0.906 & 559 \\
Portability & 0.000 & 0.000 & 0.000 & 68 \\
Performance & 0.671 & 0.905 & 0.771 & 1369 \\
Reliability & 0.821 & 0.294 & 0.433 & 187 \\
Security & 0.931 & 0.696 & 0.797 & 3161 \\
Usability & 0.616 & 0.740 & 0.672 & 1584 \\
Compatibility & 0.336 & 0.449 & 0.384 & 379 \\
\hline
Accuracy & \multicolumn{4}{c}{0.734} \\
Macro Avg & 0.607 & 0.575 & 0.566 & 7307 \\
Weighted Avg & 0.767 & 0.734 & 0.735 & 7307 \\
\hline
\end{tabular}

\label{table:multi_class_top_repo}
\end{table}

\subsection{RQ3: Comparison between QualiTagger and LLMs}

Large Language Models (LLMs) like GPT4 and smaller, more efficient models like DistilRoBERTa represent distinct trade-offs in Natural Language Processing (NLP) tasks. This research question aims to compare the performance of these two models specifically for qualities classification. The motivation behind this comparison is to understand the balance between the computational efficiency offered by DistilRoBERTa and the comprehensive capabilities provided by GPT4. By examining these trade-offs, this study attempts to identify the optimal model for practical applications, taking into account factors such as accuracy, computational cost, and resource requirements. Our results clarify which model is more suitable for the classification of qualities in issue trackers.

Our study compared the performance of our various quality fine-tuned  DistilRoBERTa  model against the out-of-the-box GPT4o (“o” for “omni”) model on the same test set, revealing significant improvements across all metrics as shown in Table \ref{tab:GPT4 _performance_metrics}. For most of the qualities, DistilRoBERTa outperformed GPT4o in precision and recall. While Portability, Security and Performance  report similar results, all the other qualities are better classified by DistilRoBERTa. The results demonstrate the superior performance of our fine-tuned DistilRoBERTa  model across all key metrics. This comparison underscores the effectiveness of fine-tuning in enhancing model performance and highlights the potential of the smaller, more efficient DistilRoBERTa  model in achieving high accuracy in quality classification tasks.

\begin{table}[h]
\centering
\caption{GPT4o Evaluation Metrics Categories}
\label{tab:GPT4 _performance_metrics}
\begin{tabular}{lrrrrrr}
\toprule
       Category &  Precision &  Recall &  Accuracy &    MCC &  F1-Score &   AUC \\
\midrule
    Portability &      0.845 &   0.620 &     0.753 &  0.526 &     0.715 & 0.753 \\
Maintainability &      0.600 &   0.480 &     0.580 &  0.163 &     0.533 & 0.580 \\
    Reliability &      0.638 &   0.880 &     0.690 &  0.411 &     0.739 & 0.690 \\
  Compatibility &      0.765 &   0.827 &     0.787 &  0.575 &     0.795 & 0.787 \\
      Usability &      0.642 &   0.907 &     0.700 &  0.439 &     0.751 & 0.700 \\
       Security &      0.930 &   0.973 &     0.950 &  0.901 &     0.951 & 0.950 \\
    
    Performance &      0.930 &   0.880 &     0.907 &  0.814 &     0.904 & 0.907 \\
\end{tabular}
\end{table}

\paragraph{\textbf{Statistical Testing for QualiTagger vs GPT 4o:}}

%\subsection{RQ4: Identification of top qualities}

\begin{table}[h!]
\centering
\caption{Comparison of GPT-4o (Model 1) and QualiTagger (Model 2)}
\label{table:model_comparison}
\small
\begin{tabular}{
    l
    S[table-format=1.4]
    c
    c
    c
    c
}
\toprule
\multirow{2}{*}{\textbf{Category}} & \multicolumn{2}{c}{\textbf{Statistical Tests}} & \multicolumn{3}{c}{\textbf{Effect Size and CIs}} \\
\cmidrule(lr){2-3} \cmidrule(lr){4-6}
& {\makecell[t]{\textbf{McNemar} \\ \textbf{(p-value)}}} & {\makecell[t]{\textbf{Bootstrap} \\ \textbf{(Sig. Diff)}}} & {\makecell[t]{\textbf{Cliff's} \\ \textbf{Delta}}} & {\makecell[t]{\textbf{Model 1} \\ \textbf{CI (95\%)}}} & {\makecell[t]{\textbf{Model 2} \\ \textbf{CI (95\%)}}} \\
\midrule
Performance    & 1.0000 & No   & -0.3536 (M) & [.71, .87] & [.74, .90] \\
Usability      & 0.5413 & No   & -0.2316 (S) & [.78, .89] & [.78, .94] \\
Compatibility  & 0.2649 & Yes* & -0.9384 (L) & [.60, .78] & [.73, .90] \\
Reliability    & 1.0000 & No   & 0.2552 (S)  & [.71, .86] & [.66, .83] \\
Portability    & 0.1496 & Yes* & -0.8232 (L) & [.55, .81] & [.74, .88] \\
Maintainability& 0.0034 & Yes* & -0.9984 (L) & [.47, .69] & [.75, .91] \\
Security       & 0.6875 & No   & 0.6928 (L)  & [.93, 1.0] & [.91, .98] \\
\bottomrule
\multicolumn{6}{l}{\small *Yes indicates Model 2 (QualiTagger) is better. S: Small, M: Medium, L: Large effect size.}
\end{tabular}
\end{table}

The statistical tests comparing QualiTagger (Model 2) and GPT-4o (Model 1) in Table \ref{table:model_comparison} reveal significant differences in their ability to identify software quality attributes. QualiTagger consistently outperforms GPT-4 in \textit{Compatibility, Portability, and Maintainability,} exhibiting large effect sizes.  Notably, both models demonstrate comparable performance in identifying Performance and Reliability attributes. While GPT-4o appears slightly better at identifying Security attributes, the difference is not statistically significant. These findings suggest QualiTagger is significantly better at identifying specific quality attributes compared to GPT-4o, particularly those related to software structure, adaptability, and maintainability.

\section{Evaluation of QualiTagger In Industry}
We evaluated the Qualitagger binary classification model on a dataset provided by Visma, a multinational (33 countries in Europe and Latin America) software company encompassing over 180 entrepreneurial software companies. Visma develops and delivers software to small businesses, medium businesses, and the public sector. Some example of software delivered by Visma are mission critical business systems, e.g. payroll, HR, ERP. Visma's unique structure, characterized by high autonomy among its constituent companies and diverse team practices, offers an ideal environment for assessing the model's ability to identify and classify quality issues across a range of project contexts. For this study, we collaborated with the Chief Software Quality Engineer and a Security Business Analyst, who have an overview of several teams at Visma. 

\subsection{Preparation and first analysis}
First, the practitioners selected a number of issues related to several projects. In particular, the practitioners selected 3 datasets: two projects ( \textit{dataset1}, 189 issues and \textit{dataset2}, 173 issues), and one additional dataset including several issues from various projects (\textit{dataset3}, 3919 issues). All these datasets were selected for having been tagged manually by developers as "NFR" (being related to non functional requirements). It was therefore more likely to find additional labels related to the qualities we were studying. 

After a first analysis, however, we found that the most common label of interest for our study was "security", while other quality tags were much less present. In addition, one of the participants from the organization, had security background, and could additionally map other existing labels (for example issues automatically created from a tool) to the label "security", assuring that the ground truth was more accurate. We then decided to validate our model with this specific quality (security) in mind, based on the availability of labeled issues in the context.

Our first step was to run the "security" binary model out of the box to classify the issues and then compare with the existing ground truth (given by the label "security" manually added by the developers or mapped by our security contact from other labels).

The results were diverse, and for the security class we obtained: 
\begin{itemize}
    \item for \textit{dataset1}, we obtained a precision and recall of .58 and .14, which was quite low
    \item for \textit{dataset2}, we obtained 0.98 and 0.80 for precision and recall, which was excellent.
    \item for \textit{dataset3}, we obtained a precision and recall of 0.95 and 0.55, which was a mixed result
\end{itemize}

We then proceeded to analyze what possible reason could lead to such a variety of results. We identified a particular pattern that was recurring in the data, especially occurring in \textit{dataset1}. Such pattern was automatically generated by tools such as Aikido, Snyk, Polaris and Coverity (all tools to identify security vulnerabilities). 

We noticed that, removing the names of the tools from the issue text and a few other key words (such as "description"), the model would correctly recognize the issues as "security". This meant that such words were weighted too much by the model to classify non-security issues (the reason for this is not known).%A graphical example is reported below, where we show the words most contributing to the label "security" in the model classification by coloring such words according to their saliency.

%Karthik add here the pictures

After automatically adding the tag "security" to these issues that were generated by tools, our scores increased dramatically, as reported in the following results.

\subsection{Final results of QualiTagger application}

Table \ref{tab:set2_security} shows the results for \textit{dataset1}.  The model achieves excellent performance on this dataset, with near-perfect precision and recall for both classes.  The F1-scores are also very high (0.94 and 0.95), and the MCC is 0.94, indicating strong agreement between predicted and actual classifications.  This suggests that the model is highly effective at distinguishing between security and non-security instances in \textit{dataset1}.

Table \ref{tab:set3_security} presents the results for \textit{dataset2}.  While the model still performs reasonably well, we observe a noticeable drop in performance compared to \textit{dataset1}.  The precision for the Non-Security class is considerably lower (0.62), although the recall remains high (0.95).  Conversely, the Security class exhibits high precision (0.98) but lower recall (0.80).  This leads to lower F1-scores (0.75 and 0.88) and a reduced MCC of 0.64. This indicates that the model might be struggling with some ambiguity or overlap between the security and non-security classes in \textit{dataset2}.  Further investigation into the characteristics of this dataset is warranted.

Table \ref{tab:dataset4_metrics} displays the results for \textit{dataset3}.  The model demonstrates strong performance on this dataset as well, although the performance profile is different from \textit{dataset1}.  While both classes achieve high precision (0.97 and 0.98), the recall for the Security class is notably lower (0.77) compared to the Non-Security class (0.99).  This difference in recall is reflected in the F1-scores (0.98 and 0.86) and the MCC of 0.86.  This suggests that while the model is good at identifying security instances when it does classify them as such, it may be missing a significant portion of actual security instances in \textit{dataset3}.  This dataset may contain more subtle or complex security vulnerabilities that are harder for the model to detect.

\begin{table}[h!]
\centering
\caption{Metrics for \textit{dataset1} Security and Non-Security Classes}
\label{tab:set2_security}
\begin{tabular}{lcccc}
\toprule
\textbf{Class} & \textbf{Precision} & \textbf{Recall} & \textbf{F1-Score} & \textbf{MCC} \\
\midrule
Non-Security (0) & 0.88 & 1.00 & 0.94 & \multirow{2}{*}{0.94} \\
Security (1) & 1.00 & 0.91 & 0.95 & \\
\bottomrule
\end{tabular}
\end{table}

\begin{table}[h!]
\centering
\caption{Metrics for \textit{dataset2} Security and Non-Security Classes}
\label{tab:set3_security}
\begin{tabular}{lcccc}
\toprule
\textbf{Class} & \textbf{Precision} & \textbf{Recall} & \textbf{F1-Score} & \textbf{MCC} \\
\midrule
Non-Security (0) & 0.62 & 0.95 & 0.75 & \multirow{2}{*}{0.64} \\
Security (1) & 0.98 & 0.80 & 0.88 & \\
\bottomrule
\end{tabular}
\end{table}

\begin{table}[h!]
\centering
\caption{Classification Metrics for \textit{dataset3} Security and Non-Security Classes}
\label{tab:dataset4_metrics}
\begin{tabular}{lcccc}
\toprule
\textbf{Class} & \textbf{Precision} & \textbf{Recall} & \textbf{F1-Score} & \textbf{MCC} \\
\midrule
Non-security (0) & 0.97 & 0.99 & 0.98 & \multirow{2}{*}{0.86} \\
Security (1) & 0.98 & 0.77 & 0.86 & \\
\bottomrule
\end{tabular}
\end{table}

% \begin{table}[h!]
% \centering
% \caption{TDSet2\_Security Performance Metrics Before Polaris Change}
% \label{tab:tdset2_before_polaris}
% \begin{tabular}{lcccc}
% \toprule
% \textbf{Class} & \textbf{Precision} & \textbf{Recall} & \textbf{F1-Score} & \textbf{MCC} \\
% \midrule
% Non-security (0) & 0.41 & 0.85 & 0.55 & \multirow{2}{*}{ -0.013} \\
% Security (1) & 0.58 & 0.14 & 0.22 &  \\
% \bottomrule
% \end{tabular}
% \end{table}

All in all, the lesson learned was that the model worked quite well on manually written issues, while for automatically generated issues we could complement with a simple mapping. This can be considered a very promising first practical application of our QualiTagger. Based on these results, a larger study with the creation of a better tool integrated with Atlassian Jira and the participation of teams from multiple companies is planned for the future.

\section{Evaluation of QualiTagger with software engineering students}

In this section, we evaluate the performance and usefulness of QualiTagger via a study including 29 teams of students (on average 4 students per team) creating and self-tagging a total of 83 issues for a large course project where they develop a Kotlin app for real stakeholders. We asked the students to use QualiTagger after they created and tagged the issues and to evaluate its usefulness, and then we compared their output with the one from QualiTagger. This will show how the QualiTagger can be used in practice both in education and by practitioners. To evaluate QualiTagger with students, we set up an assignment during a project course, and then we analyzed the anonymous results. The study was structured as follows:

\begin{enumerate}
    \item First, the students participated in a lecture where the qualities were presented as part of the ISO/IEC 25010 on quality of software and systems. The slides were made available to the students.
    \item Then, the students were grouped in 29 teams and conducted their project, which consisted of developing a Kotlin app for a real stakeholders, the Norwegian Meteorology Institute. The project spanned during an entire semester, and the main goal of the course was to develop the app by using all the learning from previous courses. Part of the students are also software practitioners working in companies and taking the course on the side.
   \item Towards the end of the project, we asked the teams to create 3-4 issues related to their project and the studied qualities, and we asked them to self-tag such issues with the qualities. We chose to do this towards the end of the project so that they could create meaningful issues. After cleaning and removing issues that were not usable (e.g. because of language or because being too short), we obtained 83 issues.
   \item We then asked the teams to use the QualiTagger to suggest tags for their issues. To do so, we created an application on Hugging Face where the students could submit some text and they would receive a number of suggested tags. The assignment was given to the teams and not to single students.
   \item We asked the teams to submit both the original issues with the self-tags as well as the outcome of the QualiTagger so that we could compare the outcome.
   \item Finally, we told the students to rate usefulness of the QualiTagger for their project and usefulness of an AI companion with such tagging feature. They could answer 1-5. 
\end{enumerate}

\begin{table}[h]
\centering
\caption{Performance Metrics for Software Quality Attribute Classification (N = 83)}
\begin{tabular}{lc}
\hline
\textbf{Metric} & \textbf{Value} \\
\hline
Hamming Loss & 0.1515 \\
Micro-averaged Precision & 0.6230 \\
Micro-averaged Recall & 0.6441 \\
Micro-averaged F1 & 0.6333 \\
At Least One Match Rate & 0.7952 \\
\hline
\end{tabular}

\label{tab:classification_metrics}
\end{table}

Table \ref{tab:classification_metrics} presents the performance metrics of QualiTagger evaluated against student-generated labels across 83 data points. This comparison between our model and student assessments provides insights into the model's ability to mimic expert-level classification. 

The QualiTagger achieves a Hamming Loss of 0.1515, indicating only 15.15\% label prediction errors compared to student tags. Micro-averaged precision (0.6230) and recall (0.6441) demonstrate that the model correctly identifies 62.30\% of predicted attributes and 64.41\% of actual attributes identified by students, respectively. The balanced performance is reflected in the micro-averaged F1 score of 0.6333. Notably, the At Least One Match Rate of 0.7952 shows that in 79.52\% of cases, the model correctly identifies at least one quality attribute tagged by students, highlighting its potential for initial attribute screening. These results indicate moderate to good performance in the complex task of multi-label software quality attribute classification, with the model closely approximating student-level expertise. While there's room for improvement, particularly in precision and recall, the model shows promise for practical applications in software quality assessment, especially for suggesting potential quality attributes for further expert review.

\begin{table}[h]
\centering
\caption{Evaluation for QualiTagger from 29 teams of students}
\begin{tabular}{lcc}
\hline
\textbf{Metric} & \textbf{Score Usefulness} & \textbf{Score Companion} \\
\hline
Average & 2.7241 & 2.931 \\
Median & 3 & 3 \\
StdDev & 0.9597 & 1.307 \\
\hline
\end{tabular}

\label{tab:teams_score_eval}
\end{table}

Table \ref{tab:teams_score_eval} presents the score given by the students to the QualiTagger about its usefulness and its perceived future usefulness as an AI companion with such capability.

The average and medians around the value of 3, shows that the students in general considered the usefulness as medium. However, looking at the standard deviations (0.9 and 1.3), we can infer that the opinions were quite distributed between 2 and 4. By looking at the comments from the students, different factors influenced the lower ratings, such as: the length and clarity of the students' own written issues, the clarity of the assignment, the user interface of the tool, and the lack of an explanation for the given tags. In general, factors related to the assignment, the implementation and the students' experience writing issues. While the teams giving a higher scores could actually learn more about the qualities and could compare their own tags to get a second opinion. 

All in all, based on the combined results of the overlapping tags between students and QualiTagger and their usefulness, we can conclude that the QualiTagger was fairly accurate and provided a discrete level of usefulness. After all most of the improvements needed are more related to external factors (the assignment, the implementation and the students' experience) and not too much on the accuracy of the NLP model. Considering that the QualiTagger is still a research prototype, that the course project is limited and the students are still mostly junior practitioners, the results of using the QualiTagger for the first time in practice can be considered promising.

\section{Discussion and Use cases}\label{discussion}

Classifying qualities in issues from trackers can help teams and managers tracking and visualizing what qualities are the most affected in a project, and can help prioritizing issues based on the most important qualities affected.
We will show how to use our QualiTagger in practice to analyze projects and dataset.

For project specific repositories to monitor the Qualities issues, we demonstrated QualiTagger's practical use by analyzing two major open-source projects: Microsoft VS Code and Elastic Kibana's GitHub repositories. Our model processed issue trackers to identify quality-related concerns, revealing distinct patterns: VS Code emphasized usability (13.57\%) and reliability (10.27\%), reflecting its focus on user experience, while Kibana prioritized security (18.96\%) and usability (18.72\%), aligning with its role in data analytics. These numbers show how QualiTagger can help teams to understand their quality focus areas and guide resource allocation for quality improvements.

In addition, as use case, we have chosen to analyze Technical Debt issues using GTD dataset \citep{skryseth2023}: it is important to understand to what qualities TD is related to, and understand which qualities are the most at risk because related to sub-optimal solutions. We also analyzed issues related to several programming languages to understand if there is a relationship between the reported TD in projects written in specific languages and some of the analyzed qualities.

\subsection{Project Specific repositories to monitor the Qualities issues }

{Table \ref{tab:project_quality} provides a comprehensive view of quality characteristics across two significant open-source projects: Microsoft VS Code and Elastic Kibana. The data reveals intriguing differences in the focus areas of these projects, shedding light on their distinct priorities and development philosophies.
For Microsoft VS Code, the emphasis is clearly on user experience and cross-platform functionality. Usability leads the pack with a substantial 13.57\% of issues, followed closely by reliability at 10.27\% and compatibility at 9.19\%. This distribution aligns well with VS Code's position as a widely-used, cross-platform integrated development environment (IDE). The focus on these areas suggests that the VS Code team is committed to providing a stable, user-friendly experience across various operating systems and configurations.

In contrast, Elastic Kibana shows a markedly different set of priorities. Security takes the lead with 18.96\% of issues, closely followed by usability at 18.72\%. This security-first approach is understandable given Kibana's role in data visualization and analytics, often dealing with sensitive information in enterprise settings. The high emphasis on usability, nearly on par with security, indicates that the Kibana team is equally committed to making their powerful tool accessible and intuitive for users.}

Interestingly, both projects show relatively low percentages for security issues in VS Code (0.78\%) and compatibility issues in Kibana (5.16\%). This doesn't necessarily imply that these aspects are neglected, but rather that they might be addressed more proactively or are less frequently reported by users.

\begin{table}[h!]
\centering
\caption{Project Quality Characteristics}
\label{tab:project_quality}
\begin{tabular}{
    l
    l
    S[table-format=4.0]
    S[table-format=2.4]
}
\toprule
Project & Quality Characteristic & {Total Count} & {Relative Frequency (\%)} \\
\midrule
\multirow{7}{*}{microsoft/vscode} 
    & Performance     &  207 &  7.6723 \\
    & Maintainability &  196 &  7.2646 \\
    & Security        &   21 &  0.7784 \\
    & Compatibility   &   248 &  9.1920 \\
    & Portability     &  246 &  9.1179 \\
    & Usability       &  366 & 13.5656 \\
    & Reliability     &  277 & 10.2669 \\
\midrule
\multirow{7}{*}{elastic/kibana}
    & Performance     &  901 &  7.7002 \\
    & Maintainability & 1004 &  8.5805 \\
    & Security        & 2219 & 18.9642 \\
    & Compatibility   &   604 &  5.1620 \\
    & Portability     &  611 &  5.2218 \\
    & Usability       & 2191 & 18.7249 \\
    & Reliability     & 1017 &  8.6916 \\
\bottomrule
\end{tabular}
\end{table}

{Table \ref{tab:specificity_scores} provides a nuanced look at the relationships between different quality characteristics through specificity scores. These scores offer valuable insights into how different aspects of software quality interrelate within each project.}

Across both projects, the Performance-Reliability pair shows the highest specificity scores (0.216 for Kibana and 0.130 for VS Code). This strong correlation suggests that issues affecting performance often have implications for reliability, and vice versa. It highlights the intricate relationship between these two critical aspects of software quality.

However, security-related pairs generally show lower specificity scores in VS Code compared to Kibana. For instance, the Security-Usability pair has a score of 0.123 in Kibana but only 0.005 in VS Code. This stark difference aligns with the priorities observed in Table \ref{tab:project_quality} , reinforcing Kibana's strong focus on security across various aspects of the software.

VS Code indicated higher specificity scores for pairs involving Compatibility, such as Compatibility-Usability (0.131) and Compatibility-Portability (0.134). This underscores VS Code's emphasis on providing a consistent and user-friendly experience across different platforms and environments.

\begin{table}[h!]
\centering
\caption{Specificity Scores for Label Pairs}
\label{tab:specificity_scores}
\begin{tabular}{
    l
    S[table-format=1.3]
    S[table-format=1.3]
}
\toprule
\multirow{2}{*}{Label Pair} & \multicolumn{2}{c}{Specificity Score} \\
\cmidrule(l){2-3}
& {elastic/kibana} & {microsoft/vscode} \\
\midrule
Maintainability-Performance & 0.104 & 0.055 \\
Performance-Security        & 0.045 & 0.023 \\
Compatibility-Performance   & 0.070 & 0.038 \\ 
Performance-Portability    & 0.089 & 0.055 \\
Performance-Usability      & 0.108 & 0.062 \\
Performance-Reliability    & 0.216 & 0.130 \\
Maintainability-Security    & 0.103 & 0.021 \\
Compatibility-Maintainability & 0.130 & 0.099 \\
Maintainability-Portability & 0.127 & 0.089 \\
Maintainability-Usability  & 0.135 & 0.112 \\
Maintainability-Reliability & 0.148 & 0.084 \\
Compatibility-Security    & 0.080 & 0.012 \\ 
Portability-Security      & 0.104 & 0.013 \\
Security-Usability        & 0.123 & 0.005 \\
Reliability-Security      & 0.081 & 0.019 \\
Compatibility-Portability  & 0.101 & 0.134 \\
Compatibility-Usability    & 0.074 & 0.131 \\
Compatibility-Reliability  & 0.083 & 0.064 \\
Portability-Usability      & 0.080 & 0.081 \\
Portability-Reliability    & 0.126 & 0.079 \\
Reliability-Usability      & 0.122 & 0.080 \\
\bottomrule
\end{tabular}
\end{table}

{Table \ref{tab:peak_frequencies} presents a temporal view of quality characteristics, showing their peak occurrences across monthly, quarterly, and yearly timeframes. This data provides insights into the evolving focus of each project over time.
Usability consistently shows high peak counts across all time frames, with the highest yearly peak of 107 issues in 2022. This persistent emphasis on usability underscores its critical importance in both projects, reflecting an ongoing commitment to enhancing user experience.}

{Security, while a top priority for Kibana as seen in Table \ref{tab:project_quality} , shows relatively low peak counts in this table. The highest yearly peak for security issues is only 6 in 2022. This could indicate that security issues, when identified, are addressed promptly and may not accumulate in large numbers over time.}

Interestingly, the yearly peaks for most characteristics occur in 2022. This could suggest several possibilities: an increase in development activity, improved issue tracking and reporting mechanisms, or perhaps a growing user base leading to more diverse feedback.

\begin{table}[h!]
\centering
\caption{Peak Frequencies of Quality Characteristics}
\label{tab:peak_frequencies}
\begin{tabular}{
    l
    c
    S[table-format=2.0]
    c
    S[table-format=2.0]
    S[table-format=4.0]
    S[table-format=3.0]
}
\toprule
\multirow{2}{*}{Label} & \multicolumn{2}{c}{Monthly Peak} & \multicolumn{2}{c}{Quarterly Peak} & \multicolumn{2}{c}{Yearly Peak} \\
\cmidrule(lr){2-3} \cmidrule(lr){4-5} \cmidrule(lr){6-7}
& {Year-Month} & {Count} & {Year-Q} & {Count} & {Year} & {Count} \\
\midrule
Usability      & 2019-10 & 23 & 2022-Q4 & 41 & 2022 & 107 \\
Performance    & 2020-01 & 12 & 2020-Q1 & 20 & 2022 &  49 \\
Portability    & 2021-02 & 11 & 2023-Q1 & 26 & 2022 &  55 \\
Maintainability& 2019-10 & 13 & 2019-Q4 & 16 & 2022 &  44 \\
Reliability    & 2022-08 & 14 & 2023-Q1 & 29 & 2022 &  71 \\
Security       & 2019-10 &  2 & 2022-Q3 &  3 & 2022 &   6 \\
Compatibility   & 2019-10 &  15 & 2019-Q4 &  30 & 2022 &   53 \\
\bottomrule
\end{tabular}
\end{table}

In industry settings, Qualitagger presents numerous opportunities for enhancing software development processes. Project managers can leverage this tool to gain a bird's-eye view of quality-related issues in their projects. By automatically classifying and quantifying different quality aspects, Qualitagger enables managers to make data-driven decisions about resource allocation and priority setting.

Development teams can use Qualitagger to track the evolution of quality characteristics over time. This longitudinal view can help identify trends, such as recurring problem areas or improvements in specific quality attributes. Such insights can inform sprint planning, code review processes, and overall development strategies.

Quality assurance teams stand to benefit significantly from Qualitagger. The tool can help ensure a balanced approach to software quality, preventing over-focus on one area at the expense of others. QA teams can use the data to design more comprehensive testing strategies that cover all critical quality attributes.

In the research domain, Qualitagger opens up exciting possibilities for large-scale studies on software quality trends. Researchers can apply the tool to diverse projects across various domains, programming languages, and development methodologies. This could lead to new insights into how different factors influence software quality.

Comparative analyses become more feasible with Qualitagger. Researchers can easily compare quality profiles across different software ecosystems, potentially uncovering patterns that distinguish successful projects from less successful ones.

Furthermore, Qualitagger's ability to identify correlations between different quality attributes could lead to groundbreaking insights in software engineering practices. Researchers could investigate how improvements in one quality aspect affect others, potentially informing new best practices in software development.

By providing automated, consistent classification of quality-related issues, Qualitagger bridges the gap between raw data and actionable insights. It empowers both industry professionals and researchers to delve deeper into the dynamics of software quality, ultimately contributing to the advancement of software engineering practices and the improvement of software products.
 
\subsection{Qualities affected by TD in software projects}

{Using the QualitTagger model, we labeled quality issues within GitHub issues from the Ground Truth Technical Debt (GTD) \citep{skryseth2023} Dataset, which includes information on numerous software repositories and their primary programming languages. This dataset specifically focuses on self-admitted technical debt, providing valuable insights into developers' perspectives on quality issues.

To identify these quality concerns, we used the QualiTagger to classify GitHub issues in GTD Dataset. This classification served as the foundation for our cross-language analysis, enabling us to compare the prevalence of different quality concerns across various programming languages.
The results of this analysis are presented in the accompanying Table \ref{tab:td_impact}, highlighting the unique quality challenges faced by developers working with different programming languages. This information can be valuable for researchers and practitioners seeking to understand and address language-specific quality concerns.

As clearly shown in Table \ref{tab:td_impact}, maintainability and reliability emerge as the software qualities most significantly impacted by Technical Debt (TD), together representing nearly 41.5\% of all identified TD issues. This underscores the tendency for TD to manifest in code that is challenging to modify and susceptible to errors.
Furthermore, performance, usability, and comprehensibility are moderately affected, highlighting TD's broader impact on software efficiency, user-friendliness, and understandability. Interestingly, portability appears to be the least affected quality, suggesting that TD's influence on software transferability across environments might be relatively less pronounced.}

This analysis emphasizes the pervasive nature of Technical Debt and its detrimental effects on various software quality attributes, particularly maintainability and reliability. Proactive management of TD is therefore essential to safeguard the long-term health and success of software projects.
\begin{table}[h!]
\centering
\caption{Impact of Technical Debt on Software Qualities}
\label{tab:td_impact}
\begin{tabular}{lrr}
\toprule
Quality & Total Count &  Percentage \\
\midrule
Maintainability & 13,581 & 20.80\% \\
Reliability  & 13,555 & 20.76\% \\
Performance  & 10,550 & 16.16\% \\
Usability & 10,531 & 16.13\% \\
Compatibility & 9,249  & 14.17\% \\
Security  & 5,568  & 8.53\%  \\
Portability & 2,251  & 3.45\%  \\
\bottomrule
\end{tabular}
\end{table}

\subsection{Most frequent qualities in programming languages}

{As an additional showcase, we show a cross-language analysis examining how software quality concerns expressed in GitHub issues differ among various programming languages (applied to GTD \citep{skryseth2023} datasets and not specific projects). The QualitTagger model's classifications provide the initial data, indicating which quality attributes are associated with issues in projects using different languages.

 For instance, if security-related issues are significantly more common in issues from C++ projects compared to Python projects, this could indicate intrinsic language-related factors that influence the manifestation of technical debt.
 
A deeper analysis might involve examining whether the language's paradigm (object-oriented, functional, etc.), its ecosystem, or typical use cases contribute to these differences. The implications here are profound; such insights could guide decisions on language selection for new projects, inform educational priorities for software developers, and steer language-specific improvements in software quality tools and processes.

In table \ref{tab:positive_labels_lang} we report the Count of concerning qualities for the six language identifed using QualitTagger in GTD datset \citep{skryseth2023}, which allows some comparison across languages.}

\begin{table}[h!]
\centering
\setlength{\belowcaptionskip}{10pt}
\caption{Count of  specific qualities by Programming Language identified by QualitTagger}
\label{tab:positive_labels_lang}
\begin{tabular}{lrrrrrrr}
\toprule
Language &  comp &  main &  perf &  port &  reli &  secu &  usab \\
\midrule

Go         & 1262 & 2105 & 1741 & 527  & 2528 & 902  & 1800 \\
HTML       & 340  & 787  & 511  & 74   & 818  & 573  & 904  \\
Java       & 1119 & 1232 & 1138 & 268  & 1395 & 382  & 1103 \\
JavaScript & 1511 & 2011 & 1629 & 213  & 1717 & 661  & 1528 \\
Python     & 2322 & 4388 & 2620 & 600  & 3865 & 2201 & 2372 \\
TypeScript & 2695 & 3058 & 2911 & 569  & 3232 & 849  & 2824 \\

\bottomrule
\end{tabular}
\end{table}

\subsection{Implications for research and practitioners}

In our results and discussions, we have first evaluated the performance of QualiTagger both on the test split and on OOD projects, to understand its generalizability. Then, we reported an evaluation of QualiTagger being used by students in a project, and its performance with industry self-tagged issues. Finally, we have shown how QualiTagger can be used to analyze projects and larger datasets, not tagged, to gain insights. In particular, we could analyze if Technical Debt is more prominent in some projects or in relation to specific programming languages.

For practice, we can see from the performance analysis, that QualiTagger reaches an excellent performance, minimizing both false positives and false negatives. This is key to be used as a reliable instrument to (semi-)automatically tag issues in a backlog. 

When applied QualiTagger in industry on more than 4000 issues, the classifier performed quite well. However, we had to add a manual rule to cover issues generated by tools that were not covered by the classifier. We don't see this as a very problematic issue, as this can be fixed automatically by either fine-tuning the model to the company's data, by manually adjusting the weights of the model, or by manually adding a simple automatic rule to tag the outcome of a specific tool (as we have done int this case). 

The results from the student project highlight how, in practice, a first implementation of a prototype using QualiTagger, received a fairly positive feedback, although with mixed opinions. The feedback highlighted how QualitTagger, to be used in practice, needs to be well supported by the right implementation and UI, to be complemented with useful explanations and to be used with issues that are written with a certain quality. All in all, we can conclude that the first attempt was promising. We plan to create a more sophisticated version of QualiTagger to be effectively embedded both in education but also in practice to be used by developers to get aid into (semi-) automatically tag issues in a backlog or issue tracker.

A key result to notice, is that QualiTagger works exceptionally well in recognizing qualities in OOD (Out of Distribution) projects. This implies that our model can be used out-of-the-box to classify issues. This is not possible with other similar approaches covering other concepts to be classified, for example Technical Debt: in \citep{skryseth2023}, it was evident how fine-tuning the model with additional data from the project, would be necessary to reach an acceptable performance. However, that means that the tagged project need to have at least some issues that are already labelled by the teams. This is not the case for QualiTagger, which does not need to learn about the new context to be able to classify qualities. This is an advantage for the practical applicability of QualiTagger.

As for research, our large dataset  QualiDatSet of tagged issues can be used to train novel and more optimized ML models to increase the performance of quality tagging. QualiTagger can be used, as shown with project and dataset tagging, to reliably and automatically tag issues to perform large case-studies about qualities. There will be no constraints as to whether the projects contains tagged issues, because QualiTagger would help categorize issues and would allow researchers to select the best and most fitting projects for the study. Finally, QualiTagger can be enhanced to include more qualities that can be important for a given study or organization. Having an ensemble of binary models has the advantage, with respect to the multiclass, that makes QualiTagger modular, as it's easy to add just one more binary model covering a new quality without re-training everything else.

\section{Threats to validity}\label{threats}
In this section, we discuss our limitations and threats to validity.

\textbf{Internal validity:} Our experiment proves the benefits of NLP techniques to classify text according to different software qualities. However, we acknowledge that certain text snippets are too short to find out which are the underlying qualities affected as our algorithm compares similar contexts to assign the qualities in these cases. We can mitigate this threat setting a threshold for the minimum length of the text to be analyzed (future work). Additionally, we often regard as TD issues affecting a quality: although hypothetically the relationship could go both ways (a quality issue is causing TD), in most analyzed cases a design issue is the root cause for a compromised quality. However, more research is needed to assess that this relationship is not bidirectional.

\textbf{External validity:} discusses the generalizability of the results. Our solution seem to work for the text issues related to repositories that have not been used to train the models. However, more datasets in real world organizational settings should also be studied to further assess the generalizability of our results.  

\textbf{Construct validity:} the main threat here is the reliance on the quality and diversity of the training data. Here we rely on issues that are self-classified by developers, and can be prone to mistakes. Enhancing construct validity would involve ensuring a comprehensive and representative training dataset and standardizing definitions of software quality attributes for consistent interpretation and classification. Also, double-checking the quality of the results with additional human feedback is also advisable in future work.

\section{Conclusion}\label{conclusion}

Tagging qualities in issues from issue trackers is something useful that increases visibility and provides insights, but which is generally burdensome for software teams. 

In this paper, we curate and release QualiDataSet, dataset of issues tagged with seven specific qualities, which can be used to train ML models able to automatically recognize if an issue is related to a certain quality. 

Then, we created QualiTagger, consisting of an ensemble of binary ML models, that is able to classify issues from issue trackers. Our evaluation showed that QualiTagger has an excellent performance in classifying qualities, including for OOD projects, which implies a high generalizability and applicability in practice as out of the box tool for issues classification. We also tested, we very good results, the tagger in industry context for the security quality.

Our first prototype based on QualiTagger, used with educational purposes with 29 teams of students, showed promising results. Finally, our demonstrative analysis of projects and datasets using QualiTagger, showed how to provide useful insights to software teams, managers and for research purposes. 

%\begin{appendices}

%\section{Section title of first appendix}\label{secA1}

%\end{appendices}
\bibliography{sn-bibliography}
\end{document}